\documentclass[conference]{IEEEtran}
\IEEEoverridecommandlockouts
\usepackage{cite}
\usepackage{amsmath,amssymb,amsfonts}
\usepackage{algorithmic}
\usepackage{graphicx}
\usepackage{textcomp}
\usepackage{xcolor}
\usepackage{hyperref}
\usepackage{multirow}
\usepackage{booktabs}

\hypersetup{
    colorlinks=true, 
    linkcolor=black,  
    citecolor=black, 
    filecolor=black, 
    urlcolor=black    
}

\def\BibTeX{{\rm B\kern-.05em{\sc i\kern-.025em b}\kern-.08em
    T\kern-.1667em\lower.7ex\hbox{E}\kern-.125emX}}
\begin{document}

\title{MPM-QIR: Measurement-Probability Matching for Quantum Image Representation and Compression via Variational Quantum Circuit
\thanks{*Corresponding Author: 2503001@narlabs.org.tw\\}
}

\author{
\IEEEauthorblockN{
    Chong-Wei Wang \IEEEauthorrefmark{1}\IEEEauthorrefmark{3},
    Mei Ian Sam \IEEEauthorrefmark{2},
    Tzu-Ling Kuo \IEEEauthorrefmark{1}\IEEEauthorrefmark{3},
    Nan-Yow Chen \IEEEauthorrefmark{3},
    Tai-Yue Li \IEEEauthorrefmark{3}*,
}
\IEEEauthorblockA{\IEEEauthorrefmark{1}Undergraduate Program in Intelligent Computing and Big Data, Chung Yuan Christian University, Taoyuan, Taiwan}
\IEEEauthorblockA{\IEEEauthorrefmark{2}Department of Physics, National Tsing Hua University, Hsinchu, Taiwan}
\IEEEauthorblockA{\IEEEauthorrefmark{3}National Center for High-performance Computing, National Institutes of Applied Research, Hsinchu, Taiwan}
}

\maketitle
\begin{abstract}
We present MPM-QIR, a variational-quantum-circuit (VQC) framework for classical image compression and representation whose core objective is to achieve equal or better reconstruction quality at a lower Parameter Compression Ratio (PCR). The method aligns a generative VQC’s measurement–probability distribution with normalized pixel intensities and learns positional information implicitly via an ordered mapping to the flattened pixel array, thereby eliminating explicit coordinate qubits and tying compression efficiency directly to circuit (ansatz) complexity. A bidirectional convolutional architecture induces long-range entanglement at shallow depth, capturing global image correlations with fewer parameters. Under a unified protocol, the approach attains PSNR $\geq$ 30 dB with lower PCR across benchmarks: MNIST 31.80 dB / SSIM 0.81 at PCR 0.69, Fashion-MNIST 31.30 dB / 0.91 at PCR 0.83, and CIFAR-10 31.56 dB / 0.97 at PCR 0.84. Overall, this compression-first design improves parameter efficiency, validates VQCs as direct and effective generative models for classical image compression, and is amenable to two-stage pipelines with classical codecs and to extensions beyond 2D imagery.
\end{abstract}


\section{Introduction} 
In recent years, the potential of quantum computing for data representation and learning has attracted widespread attention, particularly in the compression and generation of high-dimensional image data. Conventional Quantum Image Representation (QIR) models, such as the Flexible Representation of Quantum Images (FRQI)\cite{le2011flexible} and the Novel Enhanced Quantum Representation (NEQR)\cite{zhang2013neqr}, established two canonical approaches based on amplitude and basis encodings, respectively. A common feature of these methods is the allocation of explicit coordinate qubits for pixel addresses, typically denoted as $|y\rangle|x\rangle$ (or $|YX\rangle$). Although subsequent work, such as the Novel Generalized Quantum Image Representation (NGQR), has extended QIR to images of arbitrary dimensions and sizes\cite{xing2024ngqr}, these approaches still rely on explicit, position-aware encoding. As image resolution increases, the overhead of coordinate qubits and state-preparation circuits grows rapidly, becoming a major resource bottleneck for current Noisy Intermediate-Scale Quantum (NISQ) devices.

Meanwhile, the fusion of quantum circuits with machine learning has accelerated. The quantum convolutional neural network (QCNN)—initially introduced for quantum phase recognition \cite{cong2019quantum}—was soon repurposed for classical vision tasks, where variational quantum circuits (VQCs) act as “quantum convolution” layers to capture rich feature representations \cite{fan2023hybrid, gong2024quantum}. Beyond image classification and regression, a diverse array of quantum machine learning (QML)~\cite{schuld2023quantum,wossnig2021quantum,Maria2019quantum,chen2024cutnqsvmcutensornetacceleratedquantumsupport,an2025quantum} architectures has emerged—including Quantum Support Vector Machines (QSVMs) for kernel-based learning \cite{rebentrost2014quantum,havlivcek2019supervised,chen2024validating,ma2025robust,tai2022quantum}, Quantum Kernel–augmented LSTM networks for time-series forecasting \cite{hsu2025quantum} and other hybrid quantum–classical pipelines. Despite this rapid progress, to the best of our knowledge no VQC-centric generative framework has yet been rigorously benchmarked for directly representing and efficiently compressing images without resorting to explicit coordinate qubits.

For data compression, early QIR compression relied on classical pre-processing, such as Boolean expression minimization, to optimize circuit preparation for pixels sharing the same color value\cite{le2011flexible, zhang2013neqr}. These are fundamentally circuit optimization techniques, not learning-based models. Learning-based compression models include the Quantum Autoencoder (QAE), introduced by Romero et al., which is trained to compress quantum data by mapping a quantum state from a larger Hilbert space to a smaller latent one\cite{romero2017quantum}. This concept was extended to classical images with the "Quanvolutional Autoencoder," a hybrid model that uses a non-trainable quantum circuit as a feature extractor within a classical autoencoder framework, whose primary function is indeed data compression and learning latent representations\cite{orduz2025quantum}.

Our work introduces a generative VQC model for direct image representation, distinct from encoding-based QIR models. We remove the need for explicit position-encoding qubits. Instead, positional information is implicitly learned by the model as it maps its output probability distribution to the flattened pixel array. Compression is therefore achieved by optimizing an efficient ansatz to minimize the number of variational parameters required to represent the image, directly linking compression to model complexity.

The main contributions of this work are as follows:
\begin{itemize}
    \item We introduce MPM-QIR, a generative VQC framework for image representation that bypasses the need for explicit position-encoding qubits by matching measurement probabilities to normalized pixel intensities.
    \item We propose a novel bidirectional convolutional ansatz specifically designed to generate the long-range entanglement required to capture complex, global image correlations.
    \item We provide a comprehensive benchmark demonstrating that our approach achieves superior parameter efficiency, yielding higher reconstruction quality (PSNR/SSIM) with a lower Parameter Compression Ratio (PCR) compared to established QCNN and QAE models.
\end{itemize}

\section{Methodology}
\subsection{Measurement-Probabilities Matching for Image Representation}
\begin{figure}[!t]
\centering
\includegraphics[scale=0.08]{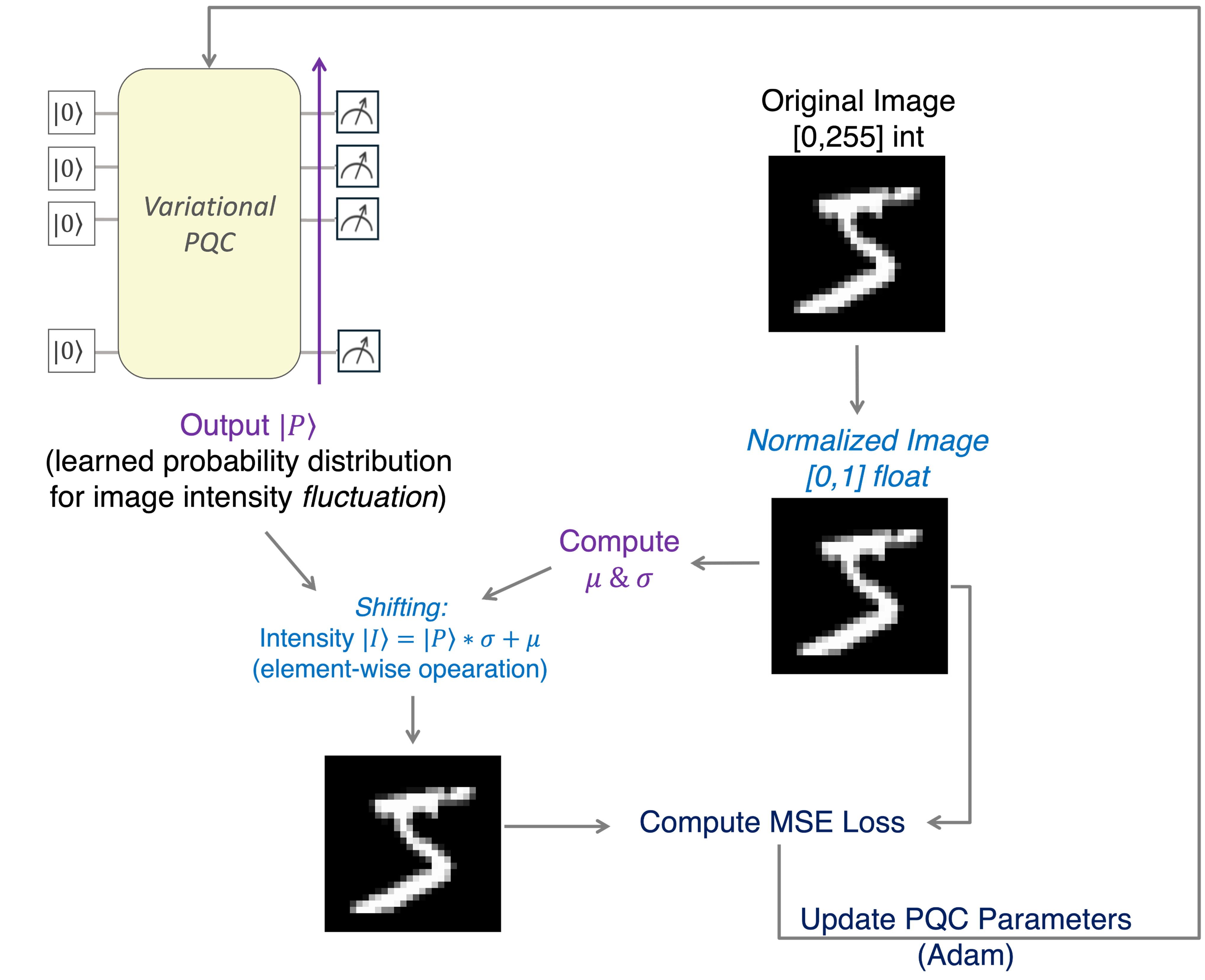}
\caption{\textbf{Overview of the MPM-QIR training pipeline.} A target grayscale image is first normalized into the range $[0, 1]$ to produce the normalized image $I_{\text{norm}}$. The VQC is initialized with the $|0\rangle^{\otimes n}$ state and executed to generate a measurement image fluctuation distribution $|P\rangle$. The output probabilities are then rescaled through a shifting operation, where parameters $\mu$ and $\sigma$ are computed from $I_{norm}$. The reconstructed image is compared with $I_{\text{norm}}$ using the MSE loss, and the VQC parameters are iteratively updated via the Adam optimizer to minimize the reconstruction error.}
\label{fig:pipeline}
\end{figure}

MPM-QIR is a generative, hybrid quantum-classical framework illustrated in Fig. \ref{fig:pipeline}. The core principle is to train a variational Quantum Circuit such that its output measurement probability distribution matches the pixel intensities of a target image after classical post-processing. As shown in the pipeline, the classical target image is first normalized into $I_{norm}$. Concurrently, the VQC, initialized to the $|0\rangle^{\otimes m}$ state, is executed to produce a probability vector. This vector undergoes classical post-processing, including rescaling and reshaping, to generate the reconstructed image, $I_{recon}$, which lies within the $[0, 1]$ interval. The VQC parameters are then iteratively optimized by minimizing the loss of the MSE between $I_{recon}$ and $I_{norm}$, while the restoration to the original $[0, 255]$ pixel range is performed only after this optimization phase. The specific designs for the Ansatz and the formal definition of the loss function are detailed in the following subsections.

\subsection{Quantum Circuit Design} 
We adapt the QCNN architecture, which uses local, translationally invariant operations to extract hierarchical features\cite{cong2019quantum}. A typical QCNN employs convolutional and pooling layers, where the pooling mechanism reduces dimensionality via mid-circuit measurements and conditional operations. However, this information-destructive pooling mechanism is unsuitable for high-fidelity generative tasks.

\begin{figure}[htbp]
\centering
\includegraphics[scale=0.28]{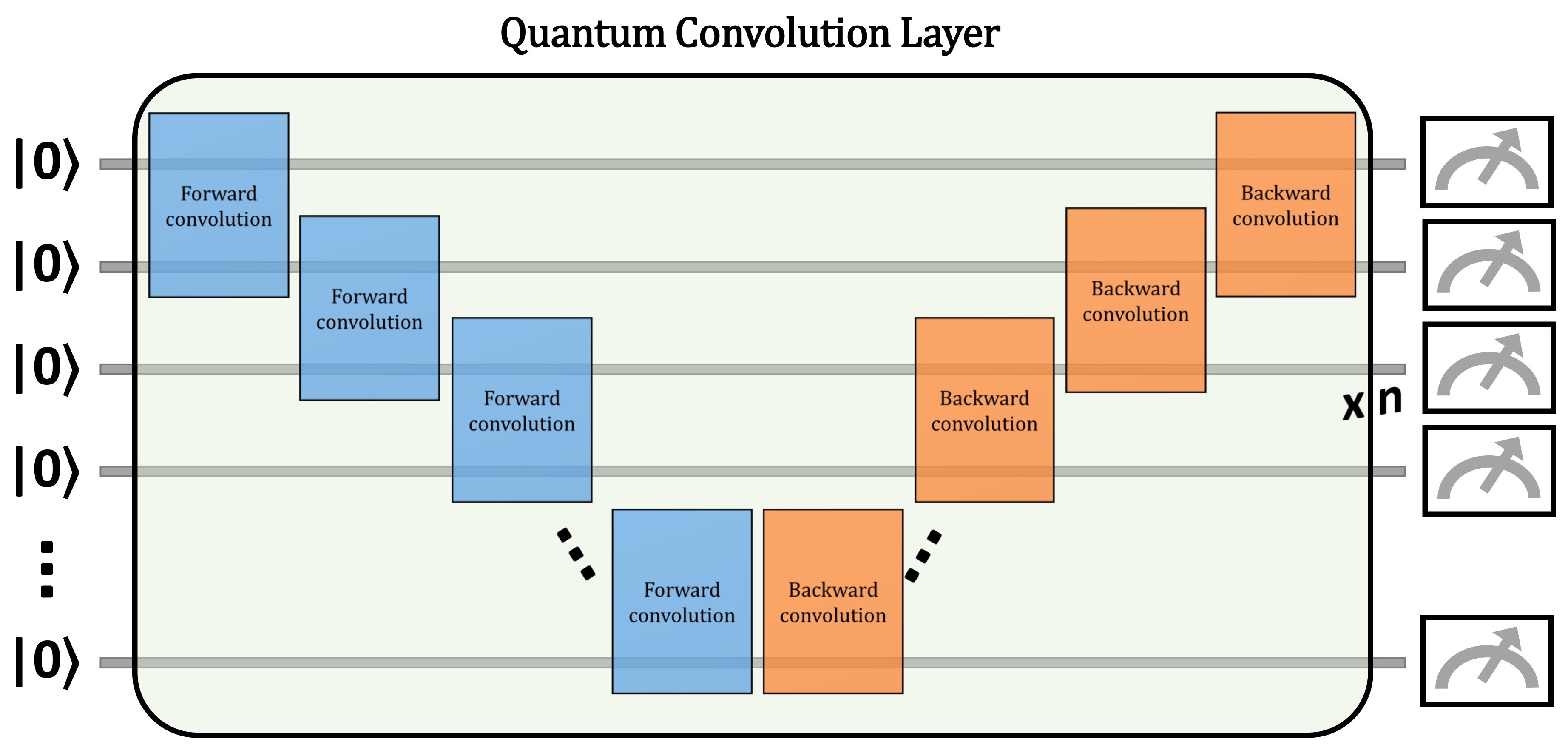}
\caption{\textbf{Schematic of the proposed bidirectional VQC architecture.}A single layer consists of a Forward Convolution pass (blue) followed by a complementary Backward Convolution pass (orange). The entire layer can be repeated ($n$ times) to increase depth.}
\label{fig:VQC}
\end{figure}

\begin{figure}[htbp]
\centering
\includegraphics[scale=0.35]{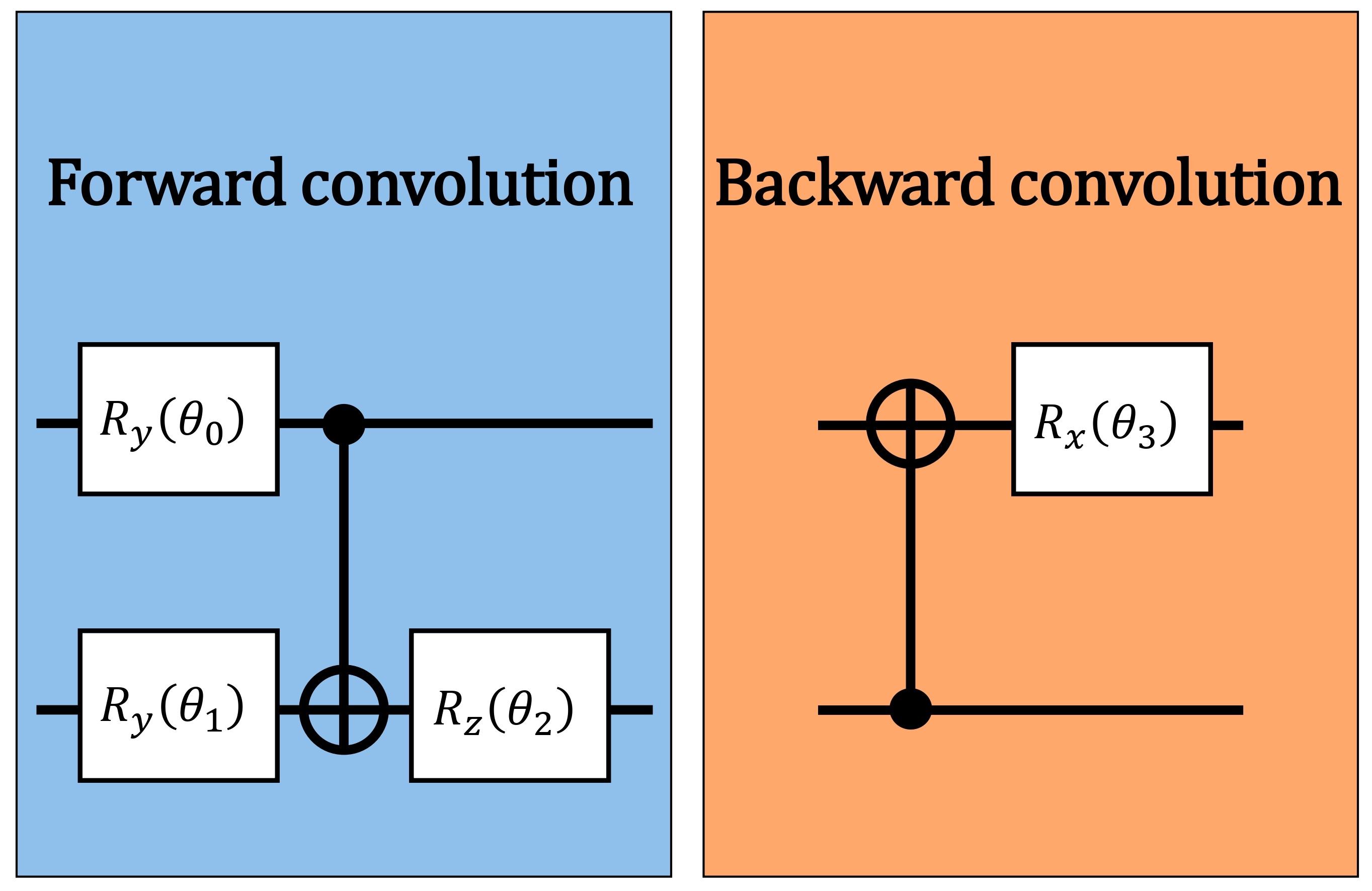}
\caption{\textbf{Gate-level decomposition of the unitary blocks.} This figure illustrates the basic structure of the quantum convolution operation used in our model. The \textit{forward convolution} (left) applies parameterized rotation gates 
$R_y(\theta_0)$, $R_y(\theta_1)$, $R_z(\theta_2)$  and controlled-NOT (CNOT) operation 
to entangle neighboring qubits and extract local correlations. The \textit{backward convolution} (right) performs the inverse entanglement operation 
using a reversed CNOT pattern and an $R_x(\theta_3)$ rotation, 
allowing information to propagate backward across qubit pairs. Together, these two blocks form a bidirectional quantum convolutional layer 
capable of encoding both forward and backward feature dependencies in the quantum state.}
\label{fig:detail}
\end{figure}

We instead introduce a novel full-width, bidirectional convolutional architecture across all qubits, depicted in Fig. \ref{fig:VQC}. Each layer is composed of a Forward Convolution Pass and an Backward Convolution Pass. The forward pass applies parameterized two-qubit unitaries (Fig. \ref{fig:detail}) in a cascading sequence across adjacent qubits, followed by the backward pass applying a complementary set of unitaries, featuring CNOT gates with swapped control and target qubits, in the reverse sequence. This dual-pass structure allows for bidirectional information flow, generating the long-range entanglement required to capture complex, global image correlations. The entire block can be repeated ($n$ times) to increase the depth and learning capacity of the circuit.

\subsection{Loss Functions} 
We employ a hybrid quantum-classical workflow. First, the target image is normalized to the range $[0, 1]$, denoted as $I_{norm}$.

The VQC, described by the unitary $U(\theta)$, where $\theta = (\theta_0, ..., \theta_{P-1})$ is the vector of trainable parameters, acts on an initial state of $m$ qubits $|0⟩^{\otimes m}$. Measurement yields a probability distribution $P_{\text{quantum}}(\theta)$, where the probability of measuring the $i$-th basis state is:
$$
P_{\text{quantum}}({\theta})_i = |\langle i | U({\theta}) | 0 \rangle^{\otimes m}|^2
$$
A reconstructed image, $I_{recon}$, is then generated via classical post-processing. This involves selecting the first $N = W \times H$ probabilities from $P_{\text{quantum}}({\theta})$, rescaling this sub-vector to match the mean ($\mu$) and standard deviation ($\sigma$) of the flattened $I_{norm}$, clipping the values to the range $[0, 1]$, and finally reshaping the vector into a $W \times H$ image.

The model is trained by minimizing the Mean Squared Error (MSE) between the $I_{norm}$ and $I_{recon}$. The loss function is defined as:
$$
L_{\text{MSE}}({\theta}) = \frac{1}{W \times H} \sum_{i=1}^{W} \sum_{j=1}^{H} [I_{norm}(i,j) - I_{recon}(i,j)]^2
$$

\subsection{Optimization Process} 
For multi-channel images, an independent VQC is trained for each color channel. The process iteratively updates the parameter vector ${\theta}$ to minimize the loss function. Gradients are computed on the quantum processor using the parameter-shift rule, and the classical Adam optimizer performs the parameter update: ${\theta}_{t+1} = \text{Adam}({\theta}_t, \nabla_{{\theta}_t} L({\theta}_t))$.

\subsection{Validation} 
We evaluate the quality of the reconstructed image against the original. For multi-channel images, the reported metrics are the average over all channels.

\subsubsection{Peak Signal-to-Noise Ratio (PSNR)}
We evaluate reconstruction quality using PSNR, defined as:
$$
\text{PSNR} = 10 \cdot \log_{10} \left( \frac{\text{MAX}_I^2}{\text{MSE}} \right)
$$
where $\text{MAX}_I=1.0$ and MSE is the Mean Squared Error.

\subsubsection{Structure Similarity Index (SSIM)}
We also measure perceptual similarity using SSIM:
$$
\text{SSIM}(x, y) = \frac{(2\mu_x\mu_y + c_1)(2\sigma_{xy} + c_2)}{(\mu_x^2 + \mu_y^2 + c_1)(\sigma_x^2 + \sigma_y^2 + c_2)}
$$
where $\mu$ and $\sigma$ represent local means and variances/covariance.

\subsubsection{Parameter Compression Ratio (PCR)}
We define PCR as the ratio of total trainable parameters to total image pixels. For an image of size $W \times H$ with $C$ channels, and a VQC with $N_{\theta}$ parameters per channel, the PCR is:
$$
\text{PCR} = \frac{\text{Total VQC Parameters}}{\text{Total Image Pixels}} = \frac{N_{\theta} \times C}{W \times H \times C} = \frac{N_{\theta}}{W \times H}
$$

\section{Result}
We evaluated the proposed circuit architecture on the MNIST, Fashion-MNIST, and CIFAR-10 datasets, comparing it against previous VQC designs under identical experimental environments, methodologies, and hyperparameter settings. Regarding the specific architectures of these benchmarks, the QCNN implementation adopts the same quantum circuit design as our Forward convolution layers. However, the standard pooling mechanism originally designed for classification is substituted with controlled-RY and controlled-RZ gates applied every two layers. This adaptation facilitates information integration similar to pooling but without reducing the dimensionality of the quantum state. While structurally distinct from the original classifier, we retain the designation "QCNN" due to these shared design principles. For the QAE baseline, we utilized the circuit architecture "B" described in \cite{romero2017quantum}. In this configuration, each layer comprises two sets of RY and RZ gates, interlaced with a sequence of $(n-1)$ controlled-RX gates to induce entanglement. 

\begin{figure}[!t]
\centering
\includegraphics[scale=0.25]{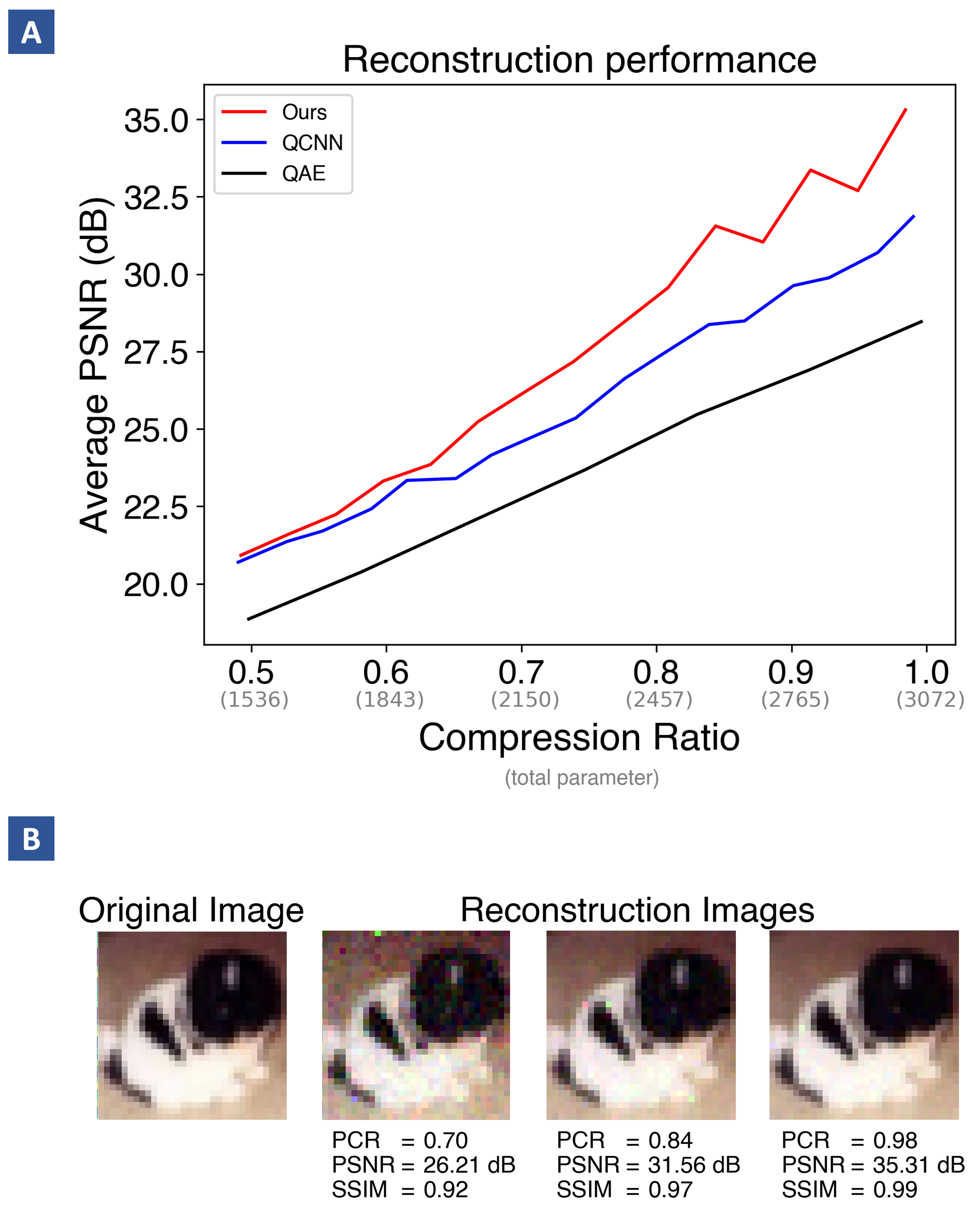}
\caption{\textbf{Reconstruction performance on CIFAR-10 dataset.} (A) Average PSNR (dB) of reconstructed images under different compression ratios for the proposed model (Ours), QCNN, and QAE. Our model consistently achieves higher reconstruction fidelity across all parameter compression levels. (B) Visual reconstruction examples from our model, along with their corresponding PSNR and SSIM scores.}
\label{fig:result}
\end{figure}

As shown in Fig. \ref{fig:result} (A), our design consistently outperforms other methods across all compression ratios, achieving a PSNR exceeding 30dB with a Parameter Compression Ratio (PCR) of 0.84.

Fig. \ref{fig:result}(B) displays visual examples of CIFAR-10 reconstructions at three different compression ratios. These examples qualitatively demonstrate that a PSNR exceeding 30dB corresponds to high perceptual quality, validating our choice of 30dB as a meaningful performance threshold.

\begin{figure}[htbp]
\centering
\includegraphics[scale=0.55]{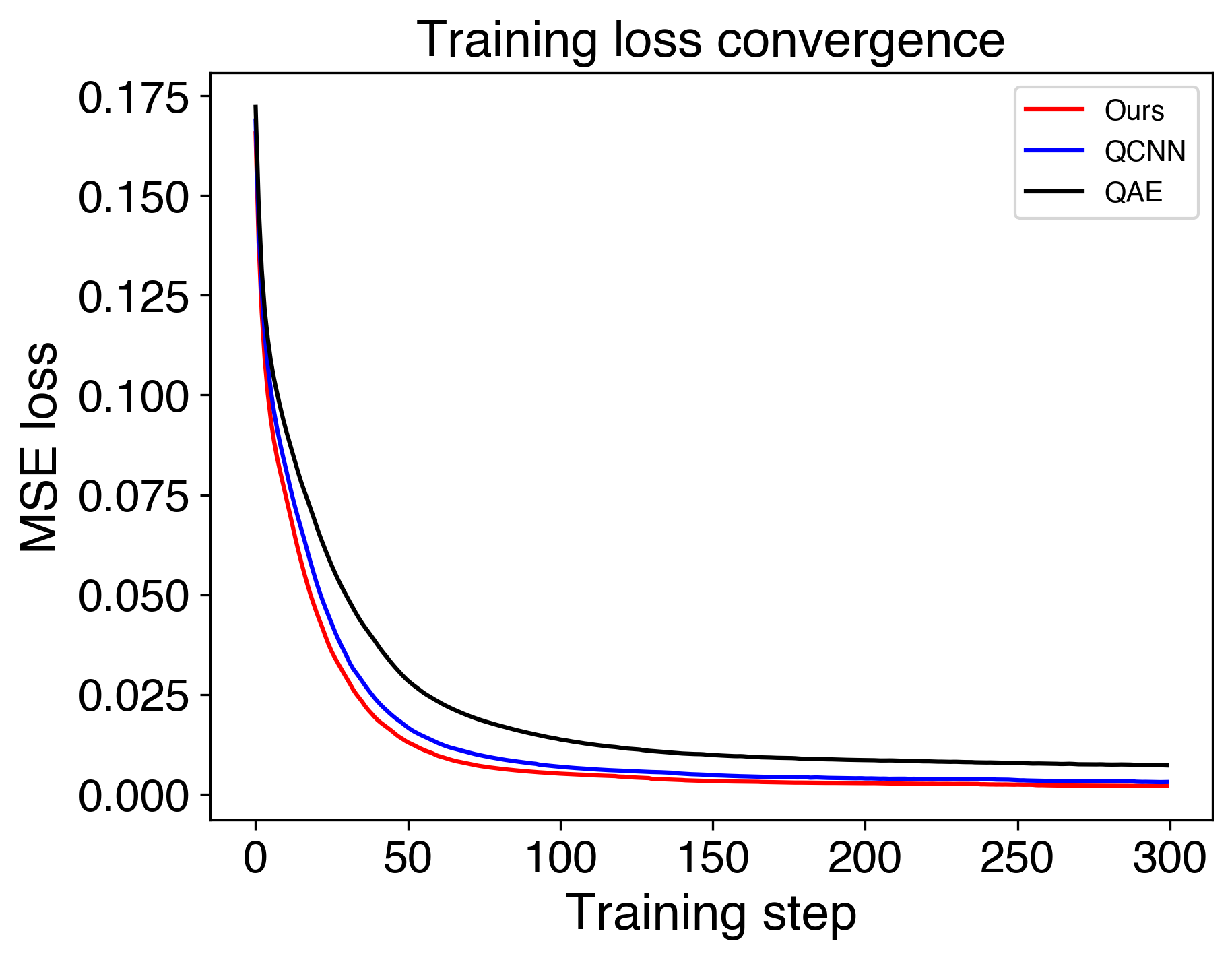}
\caption{\textbf{Convergence of training loss for different models.} The mean squared error (MSE) loss is plotted as a function of training steps 
for the proposed model (Ours), QCNN and QAE. All models show smooth convergence, but the proposed approach achieves both faster loss reduction and a lower final MSE, indicating more stable and efficient optimization.}
\label{fig:train}
\end{figure}

Furthermore, the training progression is visualized in Fig. \ref{fig:train}. Our proposed model (Ours) demonstrates a faster convergence rate and achieves a lower final MSE loss compared to the QCNN and QAE benchmarks, indicating superior optimization efficiency.

\begin{table}[htbp]
\centering
\caption{Quantitative comparison of reconstruction performance for the optimal compression model (PSNR $>$ 30dB) across the CIFAR-10, MNIST, and Fashion-MNIST datasets.}
\label{tab:compare}
\begin{tabular}{llccc}
\toprule
\textbf{Dataset} & \textbf{Method} & \textbf{PCR} & \textbf{PSNR (dB)} & \textbf{SSIM} \\ 
\midrule
\multirow{3}{*}{CIFAR-10} & QCNN\cite{cong2019quantum} & 0.87 & 28.49 & 0.95 \\
 & QAE\cite{romero2017quantum} & 0.91 & 26.91 & 0.94 \\
 & \textbf{Ours} & \textbf{0.84} & \textbf{31.56} & \textbf{0.97} \\ 
 \midrule
\multirow{3}{*}{MNIST} & QCNN\cite{cong2019quantum} & 0.69 & 28.70 & 0.77 \\
 & QAE\cite{romero2017quantum} & 0.76 & 26.72 & 0.76 \\
 & \textbf{Ours} & \textbf{0.69} & \textbf{31.80} & \textbf{0.81} \\ 
 \midrule
\multirow{3}{*}{Fashion-MNIST} & QCNN\cite{cong2019quantum} & 0.85 & 28.93 & 0.88 \\
 & QAE\cite{romero2017quantum} & 0.87 & 25.58 & 0.85 \\
 & \textbf{Ours} & \textbf{0.83} & \textbf{31.30} & \textbf{0.91} \\ 
\bottomrule
\end{tabular}
\end{table}

To evaluate the robustness and adaptability of our architecture across diverse image structures, we randomly selected five images from each of the three datasets for independent training and reported the average performance metrics. We define the "optimal result" as the model achieving a PSNR greater than 30dB with the minimum number of parameters $(N_\theta)$. This optimal result's compression ratio was then used as a baseline to select comparable results from other methods. However, since the parameter count per layer varies between architectures, a direct comparison at identical Parameter Compression Ratio (PCR) is impractical. Therefore, for this comparison, we ensured our method used the lowest parameter count $(N_\theta)$ among the compared results, placing it under a stricter compression condition.

As detailed in Table \ref{tab:compare}, across all three datasets, our proposed circuit achieves higher PSNR and SSIM values while utilizing fewer parameters (a lower PCR) than the compared methods.


\section{Conclusion}
In this study, we introduce MPM-QIR, a framework utilizing Parameterized Quantum Circuits (VQCs) for classical image representation and compression. By training the VQC, its measurement probability distribution is made to match the normalized image pixel values. Our method removes the resource-intensive position-encoding qubits found in traditional QIR models, allowing the VQC to focus on learning the correlations between pixels. Furthermore, we propose a bidirectional convolutional circuit architecture specifically designed to capture global image correlations. On the MNIST, Fashion-MNIST, and CIFAR-10 datasets, this architecture demonstrates performance superior to previous studies, capable of achieving approximately 16\% to 31\% in data savings at a PSNR$>$30dB baseline. This work validates the effectiveness of VQCs for classical image compression.

A significant advantage of this framework is its adaptability to compression tasks across various data modalities, not limited to 2D images. It can also be combined with classical compression methods to achieve a two-stage compression effect, which is crucial for long-distance transmission of large-scale data. However, as the quantum circuit architecture proposed in this study was designed specifically for image characteristics, future extensions to other tasks will likely require the exploration of alternative circuit architectures with stronger expressive power.

\section*{Acknowledgment}
The successful completion of this research was made possible by the academic resources and advanced research infrastructure provided by the National Center for High-Performance Computing, National Institutes of Applied Research (NIAR), Taiwan. We gratefully acknowledge their invaluable support.
\bibliographystyle{siamurl}
\bibliography{references}

\end{document}